\documentclass[english]{article}
\usepackage[T1]{fontenc}
\usepackage[latin9]{inputenc}
\usepackage{units}
\usepackage{amsbsy}
\usepackage{amstext}
\usepackage{amssymb}
\usepackage{esint}

\makeatletter

\newcommand{\lyxmathsym}[1]{\ifmmode\begingroup\def\b@ld{bold}
  \text{\ifx\math@version\b@ld\bfseries\fi#1}\endgroup\else#1\fi}

\newcommand{\binom}[2]{{#1 \choose #2}}

\@ifundefined{date}{}{\date{}}
\makeatother

\usepackage{babel}
\begin{document}

\title{Theoretical Bitcoin Attacks with less than Half of the Computational
Power (draft)}

\author{Lear Bahack%
\thanks{A math student at the Open University of Israel. Email: lear.bahack@gmail.com%
}}
\maketitle
\begin{abstract}
A widespread security claim of the Bitcoin system, presented in the
original Bitcoin white-paper, states that the security of the system
is guaranteed as long as there is no attacker in possession of half
or more of the total computational power used to maintain the system.
This claim, however, is proved based on theoretically flawed assumptions. 

In the paper we analyze two kinds of attacks based on two theoretical
flaws: the Block Discarding Attack and the Difficulty Raising Attack.
We argue that the current theoretical limit of attacker's fraction
of total computational power essential for the security of the system
is in a sense not $\frac{1}{2}$ but a bit less than $\frac{1}{4}$,
and outline proposals for protocol change that can raise this limit
to be as close to $\frac{1}{2}$ as we want. 

The basic idea of the Block Discarding Attack has been noted as early
as 2010, and lately was independently though-of and analyzed by both
author of this paper and authors of a most recently pre-print published
paper. We thus focus on the major differences of our analysis, and
try to explain the unfortunate surprising coincidence. To the best
of our knowledge, the second attack is presented here for the first
time. 
\end{abstract}

\section{The Block Discarding Attack}

The attack is based on (or can be much amplified by) the assumption
that the attacker can achieve \textquotedbl{}Network Superiority\textquotedbl{}
by maintaining many direct network connections, much above the average
of a single user. As explained in the previous section, when two blocks
are released around the same time, the one that will be propagated
faster has much higher chance to be eventually confirmed. The ability
to make one's block be propagated much faster is part of what we regard
as network superiority, while the other part is the ability to become
instantly aware of any new released block in the network. 

Propagation of blocks is relatively slow \textendash{} the average
time it takes for a node to be informed of a new block is 12.6 seconds\cite{key-4}
\textendash{} since propagation delay composes both of the data transmissions
time and the blocks verification time (a node verifies each block
before it propagates it to its neighbors). Therefore, an attacker
that maintains many slave nodes all across the network that are programmed
to propagate her blocks without verification and to send her new received
blocks without verification, is most definitely expected to acquire
network superiority. That is, as long as the network is homogeneous,
as the distributed Bitcoin network is ideally supposed to be. Propagation
of the attacker's block can be accelerated even further by composing
empty or relatively short blocks, whose verification (by the non-slave
nodes) is faster. 

Let's assume an attacker with $0<p<\frac{1}{2}$ fraction of the total
hash power achieves total network superiority, meaning she %
\footnote{For simplicity we choose to adapt the feminine form. This was chosen
by flipping a coin.%
} is instantly informed of any new released block and her generated
blocks always win the race when they are released at the same time
as a competitive block. Then the attacker will lose nothing by secretly
holding each new generated block until a competitor is found and then
release it immediately, and while holding the block treating it like
it was already accepted into the chain, i.e. mining the next block
on top of the temporary-secret block. 

Normally, when the attacker generates $x$ blocks and the rest of
the network generates $y$ blocks, and each one of the blocks is mined
on top of the previous generated one, the chain eventually grows by
$x+y$ more blocks. However in time of attack, if the attacker generates
$x$ blocks and the rest $y$ blocks, then all of the attacker's blocks
will eventually get into the chain while only $y-x$ of the other
blocks will get into the chain, so the chain eventually grows by only
$y$ more blocks: 

Each block of the attacker is released when another block is found
and hence it is used to \textquotedbl{}replace\textquotedbl{} the
competitive block within the chain. So if the attacker mines $x$
blocks, $x$ blocks of the rest of the network will be discarded,
and replaced by the attacker's blocks. The total block-chain growing
rate will be as if the attacker doesn\textquoteright{}t mine at all,
that is $1-p$ times the normal rate. 

Difficulty adjustment then lowers the difficulty so there will be
approximately the same number of generated blocks within the same
period of time. The total share of the attacker's blocks out of the
block-chain is then raised from $p$ to $\frac{p}{1-p}$, raising
the attacker's profits.

Lows of economy dictate that the cost of hash-power invested into
mining should be around the expected reward. The expected reward of
the non-attacker miners is now only $\frac{1-2p}{1-p}$ times than
before, so the total hash-power of the honest miners is about to decline
as more miners leave the system. By essence that means the attacker's
share of the total hash-power is about to exceed p, so that the attack
becomes more efficient and hence there are more miners to leave the
system\ldots{} the process can halt on some equilibrium or continue
until all honest miners leave. 

In real-world though, the retirement of miners is expected to be less
significant, since the cost of mining is divided between the cost
of electricity and network connection, and the cost of dedicated ASIC
machinery. Leaving the system can immediately stop the wastage of
electricity, while regaining the (relatively high) cost of ASIC on
times there are less buyers and more sellers is not trivial. Assuming
the attacker isn't generously willing to buy the unwanted ASICs, honest
miners might nevertheless continue to mine. 

Hence the paper is more focused on the reasonability of each attack
(meaning, whether it is profitable or not) rather than the exact theoretical
long term outcome of a reasonable attack. Yet our theoretical equilibrium
analysis is of real-world importance: it is applicable to crypto-currencies
that are based on more economically liquid means of computation such
as Litecoin, or SHA-256 based smaller crypto-currencies, in which
an honest miner can just switch to Bitcoin in case of attack. 

To analyze the possible equilibrium, let $b$ be the hash-power of
the attacker, $g$ the initial hash-power of the honest network, and
$h>0$ the new hash-power of the honest network when a possible equilibrium
is reached. For simplicity let the hash-power unit we use be such
that $b+g=1$, or equivalently, $b=p$. 

The expected number of (eventually accepted) mined blocks per hash-power
unit of an honest miner in the equilibrium state is the same as what
the expected number of mined blocks per hash-power unit was before
the attack. Since the total hash-power of confirmed blocks in the
equilibrium state is $h$, we get $\frac{\nicefrac{g}{(b+g)}}{g}=\frac{\nicefrac{(h-b)}{h}}{h}$.
By convention $b+g=1$, so we get $h^{2}=h\lyxmathsym{\textendash}b$,
or $h=\frac{1}{2}(1+\sqrt{1-4b})$. That means the fraction of the
attacker out of the new total hash-power is $\frac{b}{h+b}=\frac{2p}{2p+1+\sqrt{1-4p}}$.

For $p=\frac{1}{4}$ that means $\frac{1}{4}$ of the initial hah-power
has left; attacker has acquired a fraction of $\frac{1}{3}$ of the
new hash-power and gets twice as much revenue as before; and the difficulty
is half than before. For $0<p<\frac{1}{4}$, both the attacker gain
more rewards and the mining is easier than before with the same factor,
which is less than twice, and retirement is less than $\frac{1}{4}$
of the initial power. For $p>\frac{1}{4}$ the equilibrium is obviously
impossible, meaning the process will not halt until all honest miners
leave the network. 

In practice, total network superiority can never be achieved. Therefore
our analysis should be based upon both the attacker's fraction of
total initial hash-power $0<p<\frac{1}{2}$, and her network superiority
measure $0<ns<1$, defined as follows: when an honest miner mines
a new block and the attacker is quickly informed of it and tries to
release a competitive block as fast as she can, $ns$ is the probability
of the event that the next block mined by the honest network on top
of either of the competitive blocks, will be on top of the attacker's
block. We should stress that it is irrelevant to this event whether
the attacker succeeds to secretly mine yet another block before the
honest network does so. 

Interestingly, the attack is reasonable even were $ns$ is explicitly
lower than $1$, and yet more surprisingly even were $ns=0$, as long
as $p>\frac{1}{3}$. However the analysis is more complex since there
are many Block Discarding Attack strategies, and for different pairs
of $p$ and $ns$ a different strategy is best suited. The strategy
we have just presented is called $st_{\infty}$, and is part of a
hierarchical family of strategies analyzed in subsection 3.1; in subsections
3.2 and 3.3 we present some improvements to the strategies of subsection
3.1; in subsection 3.4 we examine the applicability of the Block Discarding
Attack to non-Bitcoin crypto-currencies designs; finally we suggest
a simple countermeasure to all possible Block Discarding Attack strategies
on subsection 3.5.

\subsection{The $st_{k}$ family of strategies}

\begin{figure}
\caption{The $st_{k}$ strategy, $k=0,1,2,...\ ,\infty$}

\begin{description}
\item [{on}] initialization \\
go to consensus
\item [{on}] consensus \\
$gap\leftarrow0$ \\
$public\ fork\ length\leftarrow0$ \\
mine on top of the last public block
\item [{on}] attacker (you) mine a new block \\
$gap\leftarrow gap+1$ \\
if $gap=1$ and $public\ fork\ length=k$ \\
.\qquad{}release the new mined block \\
.\qquad{}go to consensus\\
else \\
.\qquad{}mine on top of the new mined block
\item [{on}] the honest network mine a new block \\
$gap\leftarrow gap-1$\\
if $gap=-1$ \\
.\qquad{}go to consensus\\
else\\
.\qquad{}release your earliest unpublished block \\
.\qquad{}$public\ fork\ length\leftarrow public\ fork\ length+1$
\\
.\qquad{}if the honest block is mined on top of attacker's block
\\
.\qquad{}\qquad{}$public\ fork\ length=1$\\
.\qquad{}if $gap=1$ and $public\ fork\ length\ge k$ \\
.\qquad{}\qquad{}releas your secret block\\
.\qquad{}\qquad{}go to consensus\\
.\qquad{}else\\
.\qquad{}\qquad{}continue to mine on top of the same block
\item [{\rule[0.5ex]{1\columnwidth}{1pt}}]~\end{description}
\end{figure}

Under the following three assumptions, all reasonable less than 50\%
of hash-power Block Discarding mining strategies are shown to be just
the $st_{k}$, $k=0,1,2,..,\infty$ family, defined in algorithm 1: 
\begin{enumerate}
\item While the attacker's strategy might affect the mining difficulty,
it affects the attacker and the honest parties the same way. 
\item The strategy is not based on any information but the current block-chain
branches. 
\item The attacker never tries to extend a branch if there is a (strictly)
longer branch. 
\end{enumerate}
Since the attacker has less than half of the total hash-power, it
follows from assumption 1 that on any possible mining strategy, eventually
will come a moment where a new honest block is published and the attacker
has no competitive block to release. Then the attacker is forced to
switch to the honest block, according to assumption 3. 

We shall call such a moment \textquotedbl{}consensus\textquotedbl{}.
Using this term we can say that a Block-Discarding strategy is a set
of rules the attacker follows, beginning on consensus, until the next
consensus is achieved. The aim of such strategy is to increase the
fraction of the attacker confirmed blocks out of all confirmed blocks,
and its means are decision as for releasing or holding a block, and
choosing on top of which block to try mining the next. 

All strategies that fork the chain into more than two branches (meaning,
the attacker is simultaneously extending at least two different branches),
or that are not deterministic given the current state of branches,
can be shown to be not optimal. Thus in any reasonable strategy there
are always up to two competitive public branches of the same length,
one is composed solely of the attacker's block and the other solely
of honest blocks, where the public attacker's branch might have a
secret extension. 

Not mining on top of the attacker own branch is never reasonable;
releasing a secret block when the honest branch gets extended can
never harm the attacker; and releasing more blocks than needed for
the attacker's public branch to surpass the honest branch by a single
block, is never wise. Therefore, a reasonable strategy is just a rule
as for the circumstances on which to release just the single block
needed to make the two public branches even again, and the circumstances
in which it is better to release a block so that the attacker's public
branch surpasses the honest branch, and thus ensure the attacker's
branch is eventually confirmed. 

While attacker with total network superiority has nothing to lose
from secretly holding each of her blocks until a competitive is found,
attacker without total superiority takes a risk whenever she has only
a single secret block ahead of her public branch and there is a competitive
equally long honest branch: 

In case the honest network mines the next block before the attacker
and it is mined on top of the attacker's public branch,  her branch
gets confirmed and there are then just two competitive single-block
new branches. On the other hand, if this honest block is mined on
top of the honest branch (the probability of that is $1-ns>0$), the
attacker releases her last secret block and the next block too happens
to be an honest block extending the honest branch, she loses her whole
branch. 

The $st_{k}$ strategy differs from the strategy of attacker with
total network superiority only in the event that the attacker has
a single secret block and the length of the public branches is of
$k$ or more blocks. On this event, the $st_{k}$ attacker would release
her secret block, obtaining a new consensus. Thus we can view the
strategy of the total network superior attacker as $st_{\infty}$,
and $st_{0}$ as the honest mining strategy. We shall start by analyzing
the profitability of the $st_{1}$ strategy, which is of crucial importance:
as we latter prove, attacker has some better strategy than $st_{0}$
iff $st_{1}$ is better than $st_{0}$. 

We denote by $ar_{k}$ and $hr_{k}$ the corresponding average rewards
the attacker and the honest network accept, where the attacker uses
the $st_{k}$ strategy (and the honest network use $st_{0}$). The
reward is measured as the miner's number of eventually confirmed blocks
between two consecutive consensuses. We denote by $e_{k}$ the probability
to obtain between two consecutive consensuses, a situation where the
attacker holds a single secret block and the two public branches are
of exactly $k$ blocks each, assuming the attacker uses $st_{k}$.
Obviously, $ar_{k}$ $hr_{k}$ and $e_{k}$ are functions of $p$
and $ns$. 

Claim 1:
\begin{enumerate}
\item $ar_{1}=\frac{p^{3}}{1-2p}+2p^{2}(2-p)+p(1-p)^{2}ns$ And $hr_{1}=1-p+p(1-p){}^{2}(2-ns)$. 
\item The $st_{1}$ attacker resulted fraction of the total rewards is $\frac{\frac{p^{3}}{1-2p}+2p^{2}(2-p)+p(1-p)^{2}ns}{\frac{p^{3}}{1-2p}+2p^{2}(2-p)+2p(1-p)^{2}+1-p}$.
\item The $st_{1}$ attack is profitable iff $p>\frac{1-ns}{3-2ns}$.
\item The resulted mining difficulty of the $st_{1}$ attack is adjusted
to be $\frac{(1-p)(ar_{1}+hr_{1})}{ar_{1}+hr_{1}-p^{2}(2-p)}$ times
the previous difficulty. Note: we don't assume the retirement of honest
miners, which could theoretically lead to a further decreased difficulty.
\end{enumerate}
Corollary: No matter how fast is the information propagation between
honest nodes of a Bitcoin network compared to the attacker's nodes,
attacker with more than $\frac{1}{3}$ of the total hash-power would
nevertheless have a reasonable strategy, even if her network superiority
is zero.

Proof: Beginning in a consensus, the process in which the next consensus
is reached where the attacker uses the $st_{1}$ strategy, can either
start by two consecutive blocks the attacker mines before all others,
or is one of four possible paths: 
\begin{enumerate}
\item The first block to be mined is honest, and then the new consensus
is reached. 
\item First mined block is of the attacker, the second is honest and the
third is of the attacker too. When this block is mined the attacker
releases it immediately and the new consensus is reached. 
\item First is the attacker block, second is an honest block, and the third
block is an honest block mined on top of the previous honest block.
When this block is mined the 2-blocks honest branch leads to the next
consensus. 
\item First is the attacker's block, second is an honest block, and third
is yet another honest block which is mined on top of the attacker's
block. The branch of the first and third blocks then leads to the
next consensus.
\end{enumerate}
The corresponding probabilities of the four special cases are: $1-p$,
$p^{2}(1-p)$, $p(1-p)^{2}(1-ns)$, $p(1-p)^{2}ns$. The corresponding
rewards of the attacker are: 0, 2, 0, 1 and of the honest network
are: 1, 0, 2, 1. 

As for a process starting by two consecutive attacker's blocks, it
will end as soon as the honest network minimizes the 2-blocks gap
to a single-block gap. The rewards outcome depends only on the number
of steps until this gap is closed, while it doesn\textquoteright{}t
matter whether an honest block is mined on top of a previous honest
block or on top of the last block the attacker released. 

The expectancy of the number of binomial random walking moves until
we first get to the point which is one step to the right, where moving
left probability is $p<\frac{1}{2}$, is known to be $\frac{1}{1-2p}$,
including the last step to the right. Thus the average number of steps
to the left until that point, is half of the total number of steps
until that point not counting the last one, that is, $(\frac{1}{1-2p}-1)\frac{1}{2}=\frac{p}{1-2p}$.
Thus the average rewards the attacker is about to get when the process
begins with two consecutive blocks of her own is $\frac{p}{1-2p}+2$. 

The probability of that is $p^{2}$, hence we get: $ar_{1}=\frac{p^{3}}{1-2p}+2p^{2}+2p^{2}(1-p)+p(1-p)^{2}ns$
And $hr_{1}=1-p+p(1-p){}^{2}(2-ns)$. Therefore the attacker resulted
fraction of the total rewards is $\frac{ar_{1}}{ar_{1}+hr_{1}}=\frac{\frac{p^{3}}{1-2p}+2p^{2}(2-p)+p(1-p)^{2}ns}{\frac{p^{3}}{1-2p}+2p^{2}(2-p)+2p(1-p)^{2}+1-p}$. 

The $st_{1}$ strategy is reasonable iff $\frac{ar_{1}}{hr_{1}}>\frac{p}{1-p}$,
or equivalently $\frac{ar_{1}(1-2p)}{p(1-p)}-\frac{hr_{1}(1-2p)}{(1-p)^{2}}>0$,
that is, $p(3-2ns)+ns-1>0$, or equivalently $p>\frac{1-ns}{3-2ns}$.

As for the difficulty adjustment, the decreasing factor is the ratio
of the average number of chain blocks between two consecutive consensuses
and the average total number of valid blocks mined between two consensuses,
including all eventually dumped blocks. 

The average number of chain blocks is obviously $ar_{1}+hr_{1}$,
while the average total number of mined blocks is $\frac{1}{1-p}$
times the average total number of honest mined blocks. The number
of chain blocks between two consecutive consensuses is the total number
of honest mined blocks or this number plus one, in case the attacker
is the composer of the new consensus block. The probability of the
latter case is $p^{2}(2-p)$, hence $ar_{1}+hr_{1}-p^{2}(2-p)$ is
the average total number of honest blocks. {[}end of proof{]}

Claim 2: Let $k\in\mathbb{N}\cup\{0\}$. 
\begin{enumerate}
\item $ar_{k+1}=a_{k}+e_{k}(\frac{p^{2}}{1-2p}+2p(2-p)+(1-p)^{2}ns-1-k(1-p)^{2}(1-ns))$
and $hr_{k+1}=hr_{k}+e_{k}(1-p)^{2}(2-ns+k(1-ns))$. 
\item If $st_{k+2}$ is more profitable than $st_{k+1}$ then $st_{k+1}$
is more profitable than $st_{k}$. 
\item There is a reasonable Block Discarding Attack strategy if and only
if $p>\frac{1-ns}{3-2ns}$, under assumptions 1,2,3.
\end{enumerate}
Proof: The $st_{k+1}$ strategy differs from $st_{k}$ only in case
a situation is obtained where both public attacker's branch and the
honest branch are of exactly $k$ blocks and the attacker holds a
single secret block on top of her public branch. The probability that
between two consecutive consensuses this situation occurs, while playing
according to $st_{k}$, is denoted by $e_{k}$. Hence $ar_{k+1}\lyxmathsym{\textendash}ar_{k}$
is of the form $e_{k}(A-(k+1))$ where $A$ is the average reward
the $st_{k+1}$ attacker gets starting in the described situation
and until a new consensus is achieved. 

A is calculated much like $ar_{1}$. In fact, this process is equivalent
to the process that starts with a situation where the attacker have
a single secret block on top of consensus and ends with the next consensus,
where the attacker plays according to $st_{1}$, with a simple twist:
in case the attacker's first block is eventually confirmed, she is
extra rewarded with $k$ more blocks. 

When the $st_{1}$ attacker starts with a single secret block and
no public blocks, the probability that eventually this block is excluded
is $(1-p)^{2}(1-ns)$, since this can only happen where the next two
mined blocks are honest blocks and the second is on top of the first.
Thus we get $A=\frac{p^{2}}{1-p}+2p(2-p)+(1-p)^{2}ns+k(1-(1-p)^{2}(1-ns))$,
as claimed. As for $hr_{k+1}$, it is similarly calculated: $hr_{k+1}-hr_{k}=e_{k}\cdot H$
where $H$ is the average reward the honest network gets from the
twisted $st_{1}$\textendash{}similar process. 

A strategy $st_{m+1}$ is more profitable than $st_{m}$ iff $\frac{ar_{m}}{hr_{m}}<\frac{ar_{m+1}}{hr_{m+1}}$,
or equivalently $\frac{ar_{m}}{hr_{m}}<\frac{ar_{m+1}-ar_{m}}{hr_{m+1}-hr_{m}}=\frac{\frac{p^{2}}{1-p}+2p(2-p)+(1-p)^{2}ns+m(1-(1-p)^{2}(1-ns))}{(1-p)^{2}(2-ns+m(1-ns))}$.
The sequence ${\frac{ar_{m+1}-ar_{m}}{hr_{m+1}-hr_{m}}}_{m}$ decreases
until all of its elements are negative, so if $st_{k+2}$ is more
profitable than $st_{k+1}$, meaning $\frac{ar_{k+1}}{hr_{k+1}}<\frac{ar_{k+2}-ar_{k+1}}{hr_{k+2}-hr_{k+1}}$
, then $\frac{ar_{k+1}}{hr_{k+1}}<\frac{ar_{k+1}-ar_{k}}{hr_{k+1}-hr_{k}}$
too. Therefore $\frac{ar_{k}}{hr_{k}}=\frac{ar_{k+1}-(ar_{k+1}-ar_{k})}{hr_{k+1}-(hr_{k+1}-hr_{k})}<\frac{ar_{k+1}}{hr_{k+1}}$,
meaning $st_{k+1}$ is more profitable than $st_{k}$. As a consequence
there is a reasonable attack, meaning a strategy that is better than
$st_{0}$, iff $st_{1}$ is better than $st_{0}$. We already know
this is equivalent to $p>\frac{1-ns}{3-2ns}$. {[}end of proof{]}

The above proof shows us how to calculate the best result attacker
with certain $0<p<\frac{1}{2}$ and $0\le ns<1$ can achieve using
one of the $st_{k}$ strategies: we can recursively calculate the
expected rewards of $st_{k}$, simply by calculating $e_{k}$ and
using the formulas for $ar_{k+1}$ and $hr_{k+1}$. The first $k$
whose consecutive is yielding a smaller reward is the best option.
If $k$ is 0 or 1 only, there are finitely many ways to obtain the
described event whose probability is $e_{k}$, hence $e_{k}$ can
be calculated simply by summing the corresponding probabilities. For
calculating $e_{k}$ for $k\geq2$, one can use basic theory of Markov
processes. 

In case $p$ and $ns$ are such that the best strategy is of relatively
high $k$, the difference between $st_{k}$ and $st_{\infty}$ might
be insignificant, as $e_{k}$ is negligibly small. Thus, although
the $st_{\infty}$ is never optimal for $ns<1$, we shall nevertheless
use the following claim as estimation for the $st_{k}$ outcome of
relatively high $k$. 

Claim 3: The $st_{\infty}$ attacker expected fraction of the total
rewards is - 
\[
\frac{p-(1-2p)(\frac{1}{ns}-1)\Big(1-\frac{2(1-p)}{1+\sqrt{1-4p(1-p)(1-ns)}}\Big)}{1-p}
\]

Proof: In the $st_{\infty}$ strategy, the resulted block-chain contains
exactly the same number of blocks as the number of valid mined honest
blocks. However the chain is only partially composed of those blocks,
since some of them are \textquotedbl{}replaced\textquotedbl{} by the
attacker. Hence the desired fraction is the ratio between the number
of the attacker mined blocks that are eventually included in the chain
and the total number of honest mined blocks, either eventually included
or excluded. Equivalently, that is $\frac{p}{1-p}$ times the probability
of an attacker mined block to get accepted. 

We prove the claim by showing that a fraction of $\frac{1-2p}{p}(\frac{1}{ns}-1)\Big(1-\frac{2(1-p)}{1+\sqrt{1-4p(1-p)(1-ns)}}\Big)$
out of the total attacker mined blocks are eventually excluded. This
fraction is the ratio between the average number of excluded attacker
blocks between two consecutive consensuses to, and the average total
number of blocks the attacker succeeds to mine between two consecutive
consensuses. Out of the same random-walking reasons presented in the
proof of claim 1, the average total number of attacker's block between
two consensuses is $\frac{p}{1-2p}$. 

The average number of excluded blocks is a bit more complex to compute:
the probability that the honest network will surpass the attacker
when she has exactly $k$ mined blocks since the last consensus is
$Cat_{k}p^{k}(1-p)^{k+1}$ where $Cat_{k}=\frac{1}{k+1}\binom{2k}{k}$
is the $k$th Catalan number. When this happens, for each $1\le i\le k$,
the $i$th attacker's block is excluded iff all the $k+1-i$ honest
blocks that were mined after the $i$th block was publish, were not
mined on top of the attacker's public branch. Thus the average of
excluded blocks in this case is $\sum_{i=1}^{k}(1-ns)^{i}=(\frac{1}{ns}-1)(1-(1-ns){}^{k})$. 

Since $\sum_{k=0}^{\infty}Cat_{k}x^{k}=\frac{2}{1+\sqrt{1-4x}}$,
we get $\sum_{k=0}^{\infty}Cat_{k}p^{k}(1-p)^{k+1}(\frac{1}{ns}-1)(1-(1-ns){}^{k})=(1-p)(\frac{1}{ns}-1)\Big(\sum Cat_{k}(p(1-p))^{k}-\sum Cat_{k}(p(1-p)(1-ns))^{k}\Big)=(\frac{1}{ns}-1)\Big(1-\frac{2(1-p)}{1+\sqrt{1-4p(1-p)(1-ns)}}\Big)$,
So the fraction of excluded blocks out of all the attacker's mined
blocks is $\frac{1-2p}{p}(\frac{1}{ns}-1)\Big(1-\frac{2(1-p)}{1+\sqrt{1-4p(1-p)(1-ns)}}\Big)$,
as claimed. {[}end of proof{]}

\subsection{The $sst_{k}$ family of strategies}

Although assumption 2 of the previous subsection seems to be very
reasonable at first sight, the network does supplies the attacker
with external to the block-chain important information that might
help her make more justified decisions as for holding or publishing
blocks. 

When the attacker tries to discard a new published honest block, just
after the two competitive blocks are propagated and before any new
block is mined, the honest miners are divided into those who mine
on top of the honest block and those who mine on top of the attacker's
block. On average, the hash-power of the second part is a fraction
of $ns$ out of the total honest hash power. 

However, the sophisticated attacker can use slave nodes to estimate
the current fraction of the honest network that accepts her last released
block, and act accordingly: while the $st_{k}$ attacker always releases
her secret block whenever there is a single such block and the public
branches are of $k$ blocks, the sophisticated attacker may nevertheless
holds the secret block if the current winning probability happens
to be significantly higher than $ns$. On the other hand, when the
sophisticated attacker is left with a single secret block and public
branches are shorter than $k$ blocks, she may nevertheless release
it if the winning probability happens to be small. 

The author suspects that the posterior wining probabilities tend to
be either very close to 1 or 0, due to the exponential nature of the
block propagation process in the Bitcoin network\cite{key-4}. Obviously,
the distribution of posterior winning probabilities where values are
either 1 or 0, which is equivalent to the possibility to know in advance
whether a block is going to win or lose, is the best distribution
for the attacker. The distribution where winning probability is constantly
$ns$, on the other hand, is the worst. Unfortunately we don\textquoteright{}t
know the actual distribution, which is strongly dependent on the real-world
network topology, and trying to measure it is very expensive. Hence
the attacker best possible outcome should be upper and lower bounded. 

The best $st_{k}$ strategy is in fact the best sophisticated strategy
in the case of constant winning probabilities, thus it provides us
a lower bound. As for the upper bound, we define the $sst_{k}$ family
of sophisticated strategies, which can be shown to be the only reasonable
strategies for the 1/0 binary distribution. 

The $sst_{k}$ strategy differs from $st_{k}$ only in the situation
where the attacker was left with a single secret block and the public
branches are of at least k blocks: while the $st_{k}$ attacker always
releases her last block, the $sst_{k}$ attacker does so only when
she is about to lose, and with probability of $ns$ keeps holding
the secret block. 

We note that $sst_{0}$ and $sst_{\infty}$ are correspondingly identical
to $st_{0}$ and $st_{\infty}$; claim 2 has a similar version about
the $sst_{k}$ family; the $sst_{1}$ attacker fraction of total rewards
is $\frac{p^{2}(\frac{p}{1-2p}+2)+ns\cdot p(1-2p)}{p^{2}(\frac{p}{1-2p}+2)+ns\cdot p(1-2p)+(1-p)(1-p\cdot ns)}$
; and the attacker has a reasonable strategy iff $p>\frac{1-ns}{3-2ns}$,
meaning there is no profitable sophisticated strategy if there is
no regular profitable strategy. 

The last claim can be intuitively explained by noticing that for a
pair of $p$ and $ns$ such that $st_{1}$ is of the same profitability
as $st_{0}$ (and $sst_{0}$), $sst_{1}$ is also of the same profitability:
when the $sst_{1}$ attacker holds her last secret block despite having
a non-empty public branch, in case she know her branch is going to
win the competitor branch, we may regard the situation as having a
single secret block on top of consensus. Hence the only difference
between the $st_{1}$ and the $sst_{1}$ attacker is in a sense that
in some of the cases the latter keeps a single block on top of consensus
in secret, the former choose to publish it. However publishing a single
block on top of a consensus is of the same expected profits as holding
it secret, so the $sst_{1}$ attacker achieve the same profitability
nevertheless.

\subsection{Possible countermeasures}

Considering countermeasures to the described attacks, a natural direction
is to change the process of propagating new mined blocks so that attacker
will be less likely to achieve network superiority above $\frac{1}{2}$.
More specifically, we may ask the Bitcoin users to maintain a list
of all received maximal-length branches and deliver new received blocks
to all of their neighbors. Moreover, we may ask miners to randomly
choose on top of which branch to try mining the next block, and ignore
the order in which they were received. 

This natural direction has two major problems: first, it does not
change the core of the Bitcoin protocol, but rather suggests a non-obligatory
new configuration of end users' clients. A user may choose not to
maintain a list nor deliver more than one competitive block, and nevertheless
is not about to experience any resulted problem. Though the group
of Bitcoin users and miners has a motive to implement such a change,
the individual user has no incentive at all doing so. On the contrary:
as the protocol transport volume keeps rising, users would like to
minimize the delivered data. As we already know, bad incentives make
real problems in the Bitcoin system \cite{key-5}. 

The second and more crucial problem is the essentially limited success
of this direction. Due to the theoretical inability of honest miners
to recognize which one of two competitive blocks is the attacker's,
we can doubtfully guarantee the attacker network superiority to be
less than $\frac{1}{2}$. Yet even attacker with zero superiority
needs less than half of the total hash-power to have a reasonable
attack. 

Therefore we propose a different countermeasure, by tackling the block
discarding attacks from a different angle. All attack strategies are
based on continuously forking the block-chain into relatively short
branches, hence introducing a fork-punishment rule into to core of
the protocol can make those strategies unprofitable. More specifically,
we suggest not rewarding the miner of a block that has a competitive
block of another same-length branch, despite belonging to the winner
branch. There is a variety of possible implementations of this basic
idea, yet we would like to technically specify a simple possibility: 

A pair of two same-length competitive branches, composed of N blocks
at most, shall be called \textquotedbl{}fork evidence\textquotedbl{}.
A miner may include fork evidence as part of the new block she is
trying to mine, as long as the origin of the fork is less than 100
blocks deep. When a fork evidence have been successfully included
within a confirmed block, its lucky miner is rewarded half of the
total rewards the winning branch of the fork is about to gain, excluding
blocks of a sub-branch that has already been punished before, while
the owners of those punished blocks will then be totally disrewarded. 

It should be noted that by doing so we can set the reasonability threshold
of block discarding attack to be as close to half of the hash-power
as we want, by choosing big enough N, yet in practice we see no reason
to choose N which is greater than 10. There are two disadvantages
of the proposed change: the current 100-blocks delay of the mining
rewards, could not be abolished or get shortened as it can be now,
and the variance of mining profits\cite{key-6} would slightly increase. 

The reason for not delivering all of the punished blocks rewards to
the miner who supplies the fork-evidence, is to prevent the profitability
of intentionally mining on top of previous blocks in order to get
the old rewards into possession. An even further restriction, such
as maximum reward for supplying fork-evidence, is needed if fees can
make some of the rewards twice as big as the average reward. The author
is not concerned by the limited abolishment of money this proposal
is about to cause, yet there are possible mechanisms to spread the
total rewards of the punished blocks over many miners.

\section{The Difficulty Raising Attack}

On this section we show that the fundamental security claim of Bitcoin,
presented on the original paper of S. Nakamoto\cite{key-7} based
on binomial random walk, is theoretically inaccurate. As explained
on section 2, assuming difficulty and hash-power ownerships are constant,
the probability that an attacker in possession of $0<p<1$ times the
hash-power of the other network will be able to discard a block that
has been extended by $n$ sequential blocks is $p^{n}$. Please notice
that on this section we regard $p$ as the ratio between the hash-powers
of the attacker and the honest network, rather than the fraction of
the attacker's hash-power out of the total hash-power. 

However difficulty is not constant, and can be manipulated by the
attacker. The Difficulty Raising Attack enables the attacker to discard
$n$-depth block, for any $n$ and any positive $p$, with probability
1 if she is willing to wait enough time. 

On contrast with the Block Discarding Attack, on this second attack
the attacker is trying to calculate a completely competitive block-chain
whose blocks are uncorrelated to the honest network's blocks. The
two requirements each valid block must satisfy presented on section
2, will sure be fulfilled if each block timestamp precedes its previous
block timestamp and the time stamp of the last block of the chain
will be earlier than the real time at the point of the release.

\subsection{The simplified attack}

In order to demonstrate the basic idea of the difficulty raising,
let's make a relaxation of the adjustment mechanism: we say the difficulty
of a window is different from the difficulty of the previous window
by a factor of exactly the ratio between two weeks and the timespan
of the previous window, also when this ratio is more than 4:1 or less
than 1:4. 

The (simplified) attack is launched when a new window begins. The
first block of that window is going to be the last common block of
the honest network chain and the attacker's chain: the attacker secretly
calculates 2014 blocks on top of this first block of the window, each
block's timestamp is chosen to be one second ahead of its predecessor.
An even better possibility is to declare all times as being exactly
the same as the first block of the window \textendash{} this is possible,
according to section 2. 

Let the attacker hash-power be $0<p<1$ times the power of the honest
network. The attacker is going to keep try mining two blocks \textendash{}
the last of the current window and the first of the consecutive window
\textendash{} such that the difficulties sum of the faked branch will
exceeds that of the honest network. When she fails, she tries to mine
another couple of blocks on top of the last of the 2014 blocks. The
timestamp of the first block in the couple is chosen so that the second
block will have the desired difficulty. 

Using the difficulty of the first window as our difficulty unit and
10 minutes as our time unit, on time $t$ since the attack the total
honest block difficulties sum is about $t$, hence the attacker needs
to choose the difficulty $d$ of the second window such that $2015+d>t$. 

As time proceeds, $d$ becomes much greater than 1, so the expected
time needed for mining the first block is negligible compared to the
second. When the attacker takes a new trial she choose an interval
$\Delta$ that is much bigger than $\frac{1}{p}$, which is the expected
time of calculating the first block, and set $d$ to be such that
if she manages to mine her two blocks during the interval time, her
block-chain is about to surpass the one of the honest network. Her
chances of succeeding are about $\frac{\Delta\cdot p}{t-2015}$.

The integral $\int\frac{dx}{x}$ diverges to infinity, so if the attacker
continues with the strategy long enough, eventually she will win and
be able to double-spend. Interestingly, although the probability of
success is 1, the mean time it takes is infinity. However the median
time is finite and can be approximated by $\sqrt{e}\frac{2015}{p}$
units of 10 minutes for small values of $p$.

Since the attack might take a very long time, our assumption that
the sum of difficulties of the honest block-chain on time $t$ is
about $t$, needs to be reconsidered. As time proceeds it is reasonable
to assume the technology of computation improves. If both the honest
network and the attack hash-power have been increased by the same
factor, meaning $p$ hasn't changed, the attack would actually be
easier: the necessary ratio between the current difficulty of the
honest network and the artificially chosen difficulty $d$ of the
attack is $1:t$ where the honest network difficulty remains constant,
but higher otherwise. In fact, if hash-power exponentially rises with
time, this necessary ratio \textendash{} and hence the success probability
of each interval \textendash{} approaches a positive constant as $t$
approaches infinity. Thus the attacker is about to succeed on a finite
average time.

\subsection{The non-simplified attack}

We shall now describe the non-simplified attack, in which the difficulty
can be increased by a factor of at most 4 between two consecutive
windows. For simplicity, we assume that on time $t$ since the attack,
the sum of difficulties of the honest branch is exactly t. It is sound
to assume the total hash-power is constant, as a proportional increment
of both the attacker's and the honest miners' hash-power can only
make it easier for the attacker. 

Claim 4: 

Attacker forks the chain when a new window begins by mining a competitive
block on top of the first block of the window. The declared time of
each of the attacker's blocks is chosen to be 2.5 minutes ahead of
its predecessor. If the attacker continues with this strategy long
enough, no matter how small her hash-power is, on some point her chain
will surpass the honest chain. 

In order to prove the claim we use two lemmas about sums of independent
exponentially distributed random variables. We denote by $\exp(\lambda)$
a random variable whose probability density function is $\lambda e^{-\lambda x}$. 

Lemma 1: Denote $P_{bad}=\Pr\Big(\sum_{i=1}^{2016}\exp(\lambda)\ >\frac{4032}{\lambda}\Big)$.
In other words, $P_{bad}$ is the probability of $\sum_{i=1}^{2016}\exp(\lambda)$
being more than twice its expected value. Then $P_{bad}$ is the same
extremely small positive constant for all values of $\lambda$, and
for all $k\in\mathbb{N}$: $\Pr\Big(\sum\sum_{i=1}^{2016}\exp(4^{j}\lambda)\ >\sum_{j=0}^{k}\frac{4032}{4^{j}\lambda}\Big)<P_{bad}$. 

$\ $

Lemma 2: For each $\epsilon>0$ there is $\delta>0$ such that for
any big enough $k\in\mathbb{N}$ the probability that $\sum_{j=0}^{k}\sum_{i=1}^{2016}exp(4^{k}\lambda)$
is less than $\epsilon$ times its expected value, is greater than
$\delta$.

Proof of lemma 2: Let $m$ be sufficiently large so that for any $k>m$,
the expected value of $\sum_{j=m}^{k}\sum_{i=1}^{2016}exp(4^{j}\lambda)$
is less than $\frac{\epsilon}{4}$ of the expected value of $\sum_{j=0}^{m-1}\sum_{i=1}^{2016}exp(4^{j}\lambda)$,
which we shall denote by $E$ . Then the probability that $\sum_{j=0}^{k}\sum_{i=1}^{2016}exp(4^{j}\lambda)$
is less than $\epsilon$ times its expected value, for $k>m$, is
at least the probability that $\sum_{j=m}^{k}\sum_{i=1}^{2016}exp(4^{j}\lambda)$
is smaller than $\frac{\epsilon}{2}E$ times the probability that
$\sum_{j=0}^{m-1}\sum_{i=1}^{2016}exp(4^{j}\lambda)$ is smaller than
$\frac{\epsilon}{2}E$ . While the second probability is constant,
the first is greater than $1-P_{bad}$, hence there is such a $\delta$.
{[}end of proof{]}

$\ $

Proof of claim 4: Let $\epsilon=\frac{3}{4}p$ where $p$ is the fraction
of the attacker hash-power out of the honest network hash-power, and
let $\delta$ and $m$ be as of lemma 2 with respect to this $\epsilon$.
We divide the time, which is measured in units of 10 minutes, into
consecutive intervals of lengths: $2016\sum_{j=0}^{m-1}4^{j}$, $2016\sum_{j=m}^{2m-1}4^{j}$,
$2016\sum_{j=2m}^{3m-1}4^{j}$, etc. As we show next, the conditional
probability of the attacker success in any interval, given the failure
of the attacker to surpass the honest chain during all earlier intervals,
is bigger than $\delta$, which proves the claim.

The time it takes to the attacker to compute $k\cdot m$ consecutive
windows from the beginning of the fork, is distributed like $\sum_{j=0}^{k\cdot m-1}\sum_{i=1}^{2016}exp(4^{-j}p)$.
With probability higher than $\delta$ it is computed within less
time than $\epsilon\cdot\boldsymbol{E}\Big(\sum_{j=0}^{k\cdot m-1}\sum_{i=1}^{2016}exp(4^{-j}p)\Big)=\frac{3}{4}p\sum_{j=0}^{k\cdot m-1}\sum_{i=1}^{2016}4^{j}\frac{1}{p}=\frac{3}{4}\cdot2016\sum_{j=0}^{k\cdot m-1}4^{j}\leq(1-4^{-m})2016\sum_{j=0}^{k\cdot m-1}4^{j}=2016\sum_{j=(k-1)\cdot m}^{k\cdot m-1}4^{j}$
which is exactly the length of the $k$th interval. When the attacker
goes into another interval of time she doesn't calculate the block-chain
from the beginning of the fork but continues from the point she has
gone so far during the previous intervals of time, so the probability
of success within a certain interval is greater than $\delta$. {[}end
of proof{]}

A natural question of theoretical importance is whether this counterintuitive
attack has a possible protocol countermeasure. We believe the answer
is negative: the capabilities of computational devices seem to grow
exponentially with time, though the exponential base might gradually
and unpredictably change over time. Thus any protocol mechanism of
difficulty adjustment should enable a difficulty rising that is exponential
with respect to the block-chain length. 

There is a variety of adjustment mechanisms that differ from the Bitcoin
mechanism by having different window lengths, different adjustment
restrictions than having up to $4$ times or at least as $\frac{1}{4}$
times the previous difficulty, or even having the adjustment delayed
or smoothed\cite{key-8}. Yet any mechanism enabling exponentially
rise of the difficulty is vulnerable to the attacker, no matter how
small is the exponential base of the maximal rising.

\section{Estimating the real-world threats}

Both the Block Discarding Attack and the Difficulty Raising Attack
are currently unrealistic and of theoretical importance only. However,
while the mathematical calculations convincingly show there is no
actual threat out of the second attack, the first attack threat is
limited by the structure of the current Bitcoin network: the highest
share of a solo miner out of the total hash-power, the network topology
of miners and their delays when receiving and releasing blocks.

First of all it should be noted that the Block Discarding Attack is
not applicable to pools, since the attack requires the secrecy of
the new mined blocks, which cannot be guaranteed while they are shared
with all (anonymous) pool miners. The only thing a centralized pool
can do is to withhold a new block someone in the pool has found, while
the other pool members keep mining on top of the older block, until
a competitive block of non-pool miner has been released. Then the
one or more blocks the pool has found will be released too \textendash{}
and hopefully win the race over the non-pool competitor. 

By doing so the pool's relative share out of all mining rewards is
unchanged if the pool has total network superiority, and decreased
otherwise, hence the attack is unreasonable. Nevertheless the attack
can be performed by a small well organized group of solo miners. In
case one of the attackers happens to be a centralized pool owner,
the pool might be used to amplify the attack. 

The smaller is the attacker fraction of the system total hash-power,
the closer to 1 is the minimal network superiority essential for a
reasonable attack. Assuming all current solo miners' hash-power fractions
are relatively small, the attack is enabled only if achieving almost
total network superiority is possible within the real Bitcoin network,
which is very questionable.

On theory, an ideally homogeneous decentralized Bitcoin network will
enable attacker that maintains many slave nodes to achieve network
superiority. In practice, the real network topology is definitely
not homogeneous or ideally decentralized. The author believes it is
possible to get quickly informed of any new published block, yet harder
to immediately propagate a new block. The blockchain.info website
seems to assure the first part of the claim, while trying to validate
the second part is unfortunately very costly. Despite being generally
considered a negative phenomenon, the gathering of small miners into
major pools might positively affect the difficulty to immediately
propagate new blocks: while the average time it takes for a node to
be informed of a new block is said to be 12.6 seconds, the time it
takes for a pool organizer, which naturally maintains many network
connections, is expected to be much lower. 

Assuming attacker with less than $\frac{1}{4}$ of the total hash-power
have nevertheless magically achieves total network superiority and
uses the $st_{\infty}$ strategy, the resulted new equilibrium can
hardly be considered as a real threat to the system. This attacker
will not be able to do any real harm, yet the resulted decline of
the total difficulty of the chain theoretically makes the system more
vulnerable to a second more powerful attacker.

On the purely theoretical scenario where the attacker deports all
other miners, she can harm the system by launching a DoS attack. Double-spending
attack, however, is more problematic since the moment the Block-Discarding
attacker stops mining linearly, all the ex-miners will happily start
mining again, and are expected to gain awesome rewards due to the
lower difficulty. 

Yet we believe a countermeasure should be implied. Although not very
likely, it is absolutely possible that some government offended by
the dark market uses of Bitcoins decides to launch a DoS attack against
the system.

\section{Related works}

Related works Bitcoin attacks with less than $\frac{1}{2}$ of the
computational power are by no means new. Several attacks have been
known since the early days of the system, of which one simple yet
important double spending attack goes as follows: the attacker secretly
mines blocks on top of the longest branch she knows. No matter how
small is the attacker hash-power, eventually the time would come where
the attacker secret chain is longer by 6 blocks than the honest chain.
Then the attacker may send conflicting transactions to the honest
and the secret branch, and publish the secret branch just before the
honest branch close the gap. 

As for the two attacks described in this paper, although the author
comes to all the ideas and analysis presented here independently of
any related work, there are some similar works. Some of them are as
early as 2010, while others have been published during the last few
days. The alleged coincidence can be explained by the publication
of analysis regarding the flow of information in the Bitcoin network,
e.g. \cite{key-5} and the quite recent \cite{key-5}, that inspired
the block-discarding idea.

\subsection{The Block Discarding Attack}

The basic block-discarding idea, and a strategy to secretly hold new
mined block, were explicitly described in 2010-old thread of Bitcoin
technical discussions forum\cite{key-10} including numerical results
of a simplified simulation\cite{key-11}. Despite the participation
of influential Bitcoin developers in this forum discussion, the attack
has been long forgotten, probably due to allegedly being impractical.
Surprisingly, two researchers of Cornell University have recently
and independently published a pre-print paper mathematically analyzing
the $st_{1}$ strategy, which they call \textquotedbl{}Selfish Mining\textquotedbl{}\cite{key-12}.%
\footnote{Unfortunately the paper results were misleadingly propagated via the
web and media\cite{key-13}, causing disproportionate panic among
Bitcoin users.%
} 

Apart from presenting a more comprehensive analysis of the Block Discarding
Attack and its many strategies, we suggest a different view of the
attack nature: while the pre-print paper describes the attack as being
done by a mining pool, and the resulted effect of the $st_{1}$ strategy
is said to be a transfer of small miners from honest pools to the
\textquotedbl{}selfish\textquotedbl{} pool, we stress that a Block
discarding attack can only be performed by a very powerful solo miner
and that the key point of the attack is the difficulty adjustment
which enables the solo attacker to gain higher rewards. 

Moreover, the mentioned paper claims the attack creates a process
which will not end until the attacker pool oust all other miners.
As we explain, the more likely outcome is a new equilibrium with fewer
honest miners that maintain the same profits, as the difficulty adjustment
makes mining easier. Yet another difference is about the suggested
countermeasures.

\subsection{The Difficulty Raising Attack }

The fact that there is no way to determine whether a block have been
computed on its declared time or not, which is at the base of our
Difficulty Raising Attack, have been noted before in Bitcoin discussions
forums and even been used as part of two attacks\cite{key-14,key-15}. 

While concerns of \cite{key-14} were limited to the possible vulnerability
of network time differences, the false timestamps in \cite{key-15}
are used to manipulate the mining difficulty: the attack exploit the
protocol bag stated in section 2, that the time span of a window is
calculated from its first to its last block and not from the last
block of the previous window. However, the attack requires the cooperation
of more than half of the hash-power, and aimed only to increase the
rewarding of miners. Inventors of the attack explicitly state their
opinion regarding the impossibility of manipulate the difficulty to
achieve a chain with total sum of difficulties higher than the real
amount of hash-power invested to calculate the chain.


\begin{thebibliography}{10}
\bibitem{key-4}C. Decker and R. Wattenhofer. Information Propagation
in the Bitcoin Network, 13th IEEE International Conference on Peer-to-Peer
Computing, 2013. 

\bibitem{key-5}M. Babaioff{}, S. Dobzinski, S. Oren, and A. Zohar.
On Bitcoin and Red Balloons, 13th ACM Conference on Electronic Commerce,
2012. 

\bibitem{key-6}M. Rosenfeld. Analysis of Bitcoin Pooled Mining Reward
Systems. http://arxiv.org/pdf/1112.4980.pdf.

\bibitem{key-7}S. Nakamoto. Bitcoin: A peer-to-peer electronic cash
system. 2008. http: //bitcoin.org/bitcoin.pdf. 

\bibitem{key-8}Bitcoin wiki: List of alternative crypto-currencies.
https://en.bitcoin.it/wiki/List\_of\_alternative\_cryptocurrencies.

\bibitem{key-9}Bitcoin wiki: Block timestamp. https://en.bitcoin.it/wiki/Block\_timestamp. 

\bibitem{key-10}User \textquotedbl{}BiteCoin\textquotedbl{} et al.
Mining cartel attack, Bitcoin talk forum thread, 2010. https://bitcointalk.org/index.php?topic=2227.msg30064\#msg30064. 

\bibitem{key-11}User \textquotedbl{}BiteCoin\textquotedbl{} et al.
Mining cartel attack, Bitcoin talk forum thread, 2010. https://bitcointalk.org/index.php?topic=2227.msg30083\#msg30083.

\bibitem{key-12}I. Eyal and E. Gun Sirer. Majority is not Enough:
Bitcoin Mining is Vulnerable, 2013. http://arxiv.org/pdf/1311.0243v5.pdf. 

\bibitem{key-13}I. Eyal and E. Gun Sirer. Bitcoin is Broken, \textquotedbl{}Hacking
Distributed\textquotedbl{} blog, 2013. http://hackingdistributed.com/2013/11/04/bitcoin-is-broken.

\bibitem{key-14}The Timejacking Attack, Culubas blog, 2011. http://culubas.blogspot.com.

\bibitem{key-15}User \textquotedbl{}ArtForz\textquotedbl{} et al.
The time Wrapping Attack, Bitcoin talk forum thread, 2011. https://bitcointalk.org/index.php?topic=43692.msg521772\#msg521772. 

\bibitem{key-16}M. Rosenfeld et al. Dynamic block frequency, Bitcoin
forum thread, 2012. https://bitcointalk.org/index.php?topic=79837.0;all
. \end{thebibliography}
\end{document}